\documentclass[prb,amssymb,preprint,superscriptaddress,showpacs]{revtex4}   
\usepackage{graphicx}

\begin{document}

\title{Probing the role of the barrier layer in magnetic tunnel junction transport}

\author{B. B. Nelson-Cheeseman}
\email{bbnelsonchee@berkeley.edu} \affiliation{Department of
Materials Science and Engineering, University of California,
Berkeley, California 94720, USA}
\author{R. V. Chopdekar}
\affiliation{School of Applied and Engineering Physics, Cornell
University, Ithaca, New York 14853, USA} \affiliation{Department of
Materials Science and Engineering, University of California,
Berkeley, California 94720, USA}
\author{L. M. B. Alldredge}
\affiliation{School of Applied and Engineering Physics, Cornell
University, Ithaca, New York 14853, USA} \affiliation{Department of
Materials Science and Engineering, University of California,
Berkeley, California 94720, USA}
\author{J. S. Bettinger} \affiliation{Department of Materials
Science and Engineering, University of California, Berkeley,
California 94720, USA}
\author{E. Arenholz}
\affiliation{Advanced Light Source, Lawrence Berkeley National
Laboratory, Berkeley, California 94720, USA}
\author{Y. Suzuki}
\affiliation{Department of Materials Science and Engineering,
University of California, Berkeley, California 94720, USA}

\begin{abstract}
Magnetic tunnel junctions with a ferrimagnetic barrier layer have
been studied to understand the role of the barrier layer in the
tunneling process - a factor that has been largely overlooked until
recently. Epitaxial oxide junctions of highly spin polarized
La$_{0.7}$Sr$_{0.3}$MnO$_{3}$ and Fe$_{3}$O$_{4}$ electrodes with
magnetic NiMn$_{2}$O$_{4}$ (NMO) insulating barrier layers provide a
magnetic tunnel junction system in which we can probe the effect of
the barrier by comparing junction behavior above and below the Curie
temperature of the barrier layer. When the barrier is paramagnetic,
the spin polarized transport is dominated by interface scattering
and surface spin waves; however, when the barrier is ferrimagnetic,
spin flip scattering due to spin waves within the NMO barrier
dominates the transport.
\end{abstract}

\pacs{75.47.-m,85.75.-d,72.25.Mk,75.70.Cn}

\maketitle

The magnetic tunnel junction (MTJ) is one of the most simple spin
polarized devices whose transport behavior has yet to be completely
understood. Its basic device characteristics have been explained in
terms of a simple quantum mechanical model of two spin polarized
reservoirs separated by a potential barrier over two decades
ago.\cite{Julliere75} In this picture, depending on the relative
orientation of the magnetization in the two ferromagnetic
electrodes, a high or low resistance results. However, this
simplified picture does not accurately describe the behavior of real
MTJs. Improvements in the electrode-barrier interfaces have now led
to the observation of magnetic tunnel junction behavior at room
temperature\cite{Moodera95} and the realization that the transport
through these structures is extremely sensitive to the interface
scattering and spin polarized interface density of states of the
electrodes.\cite{Woods04} It was not until recently, however, that
the importance of understanding the role of the barrier layer in the
tunneling process was recognized in experimental and theoretical
studies of MTJs with MgO barriers.\cite{Butler01} In these
junctions, a number of factors dictated by the barrier layer were
recognized to be important, including the chemical bonding between
the atoms at the electrode-barrier interface, interface resonance
states, and the symmetries of the propagating states in the
electrodes and evanescent states in the barrier layer. In order to
develop a complete picture of spin polarized transport in MTJs, the
role of the barrier layer in the tunneling process as well as
electrode-barrier interface effects must be generalized.

There have been some efforts to elucidate the role of the barrier
layer by introducing magnetism into the barrier through dilute
doping of a nonmagnetic barrier with magnetic ions or the use of
paramagnetic barriers. Jansen \emph{et al.} found that $\delta$ -
doping of Al$_{2}$O$_{3}$ barriers with Fe ions increased the
junction magnetoresistance (JMR) as a function of Fe content in the
barrier.\cite{Jansen99} The enhanced JMR was explained in terms of
spin filtering of the tunneling electrons, whereby a preferential
transmission (increased conductance) exists for one type of spin due
to the polarized barrier states.\cite{Jansen00} Other studies of
MTJs with paramagnetic barriers suggest that the presence of
paramagnetism in the barrier layer does not preclude large JMR
values or distinct switching
characteristics.\cite{Jo00,Hu02,Alldredge06} These results suggest
that magnetism in the barrier can be spin preserving rather than
adding to spin-flip scattering via spin waves. If weak magnetism in
the barrier is spin preserving, then incorporating long-range
ferromagnetic order in the barrier layer of a MTJ should provide a
system where the role of the barrier layer can be readily probed as
long as the barrier does not magnetically couple the two
ferromagnetic electrodes.

In this paper, we demonstrate magnetic tunnel junction behavior in
junctions with ferromagnetic electrodes and a barrier layer with
long range ferrimagnetic order. Comparison of junction behavior when
the barrier layer is ferrimagnetic to when it is paramagnetic
enables us to probe the role of the barrier layer in one of the
simplest spin polarized devices. In particular, we focus on
Fe$_{3}$O$_{4}$/NiMn$_{2}$O$_{4}$$($NMO$)$/LSMO
(spinel/spinel/perovskite) junctions to preserve the magnetic
decoupling of the electrodes in the presence of a magnetic barrier
layer. Magnetization measurements of the layers as well as at the
interfaces indicate that we have two ferromagnetic electrodes that
switch independently despite the presence of a
ferrimagnetic-paramagnetic barrier layer. Spin polarized transport
measurements reveal junction magnetoresistance values as high as
-30\% (normalized to 8 kOe). Two different conduction mechanisms are
observed which directly highlight the passive or active role of the
barrier layer in the spin transport. Above the T$_{C}$ of the NMO
barrier, when the barrier layer is paramagnetic, the
electrode-barrier interfaces dominate the spin transport, resulting
in an asymmetric bias dependence of the JMR and inelastic tunneling
spectra (IETS). Below the T$_{C}$ of the NMO barrier, the
ferrimagnetism in the barrier dominates the spin transport,
resulting in a transition to a symmetric bias dependence of the JMR
and IETS.

The trilayers of Fe$_{3}$O$_{4}$/NMO/LSMO were grown by pulsed laser
deposition on SrTiO$_{3}$ (STO) (110) substrates following Ref. 9.
LSMO and Fe$_{3}$O$_{4}$ were chosen for their highly spin polarized
nature as both have been theorized and shown to be half-
metallic.\cite{Hu02} Since isostructural barrier layers have been
shown to greatly increase the JMR values for epitaxial
Fe$_{3}$O$_{4}$ MTJs,\cite{Hu02} a ferrimagnetic spinel,
NiMn$_{2}$O$_{4}$, was selected. NMO films were grown at 550
$^\circ$C in 10 mTorr of a 99\% N$_{2}$/1\% O$_{2}$ gaseous mixture.
For these deposition conditions, the NMO film was insulating with an
onset of ferrimagnetism around 60 K, a large coercive field of 1.6 T
at 30 K, a magnetization of 0.8 $\mu_{B}$ per formula unit, and an
inverse spinel cation site distribution.\cite{Unpublished} Barrier
thicknesses were 20, 30, and 45{\AA}. The films grow epitaxially on
the STO substrate with excellent crystallinity as confirmed by x-ray
diffraction and Rutherford backscattering channeling measurements.
The MTJs were fabricated by conventional lithography and Ar ion
milling.

The bulk magnetism of the samples was investigated by a
superconducting quantum interference device (SQUID) magnetometer,
while the interface magnetization was investigated by X-ray magnetic
circular dichroism (XMCD) at the Advanced Light Source on BL4.0.2.
The transport measurements were conducted in magnetic fields up to
8kOe and temperatures from 15 to 300 K. The bias and temperature
dependence of the JMR were calculated in accordance with Julliere's
model by the following equation: $\frac{\triangle
R}{R(P)}$$\times100$, where
$\triangle$R=R$_{AP}$-R$_{P}$,\cite{Julliere75} and the reference
(parallel magnetization) resistance was taken at 8 kOe.

Magnetic hysteresis loops of the trilayers show distinct switching
of the two ferromagnetic layers. We observe large coercive field
differences between the LSMO and Fe$_{3}$O$_{4}$, thus creating
well-defined parallel and antiparallel magnetization states at all
temperatures. However, the magnetization of the NMO barrier cannot
be probed with bulk methods due to its small magnetic signal
compared to that of the ferromagnetic electrodes.

In order to probe the magnetic behavior at the interfaces and within
the barrier layer, element-specific XMCD was used in the total
electron yield mode. Since the total electron yield method is
surface sensitive, we studied heterostructures of LSMO (40 nm)/NMO
(5 nm) and LSMO (40 nm)/NMO (5 nm)/Fe$_{3}$O$_{4}$ (5 nm) to probe
both interfaces. Figure 1 shows the Mn XMCD spectra for the LSMO/NMO
bilayer sample for various temperatures. A pronounced valley (peak)
in the XMCD is evident at 640 eV (642.5 eV) at lower temperatures
resembling XMCD spectra of NMO single films.\cite{Unpublished} This
feature disappears above 60 K, resulting in Mn XMCD spectra at 80
and 100 K resembling that of LSMO thin films.\cite{Stadler99} In
order to probe the origins of the Mn magnetism, hysteresis loops
were taken at 640 and 642.5 eV. Hysteresis loops taken at 642.5 eV
show magnetically soft Mn with coercive fields $<$0.5 kOe over the
entire temperature range. Those taken at 640 eV show a transition
from the soft loops above 60 K (similar to those at 642.5 eV) to
magnetically harder loops that do not reach saturation by 8 kOe,
resembling SQUID hysteresis loops of single NMO films. This
transition indicates the onset of ferrimagnetism in the magnetically
hard NMO barrier at 60 K.

\begin{figure}
\center{\includegraphics[width=8 cm]{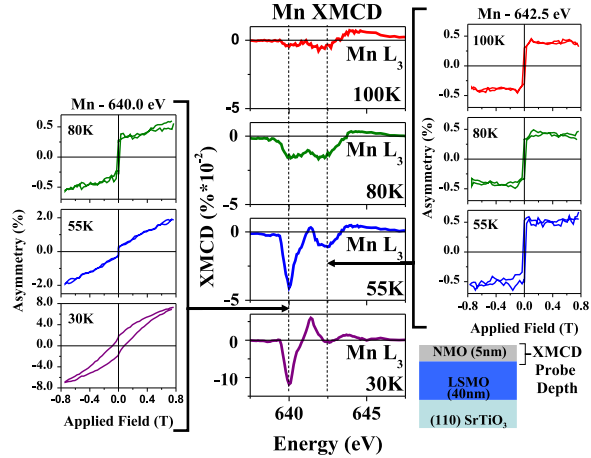}}
\caption{(Color online) Surface sensitive XMCD investigating the
LSMO-NMO interface magnetism. Normalized Mn XMCD of the NMO-LSMO
interface, and Mn hysteresis loops taken at 640.0 eV and 642.5 eV as
a function of temperature.}
\end{figure}
From the XMCD hysteresis loops in Fig. 1, it is clear that the Mn
magnetism in the NMO and LSMO is probed at 640 eV whereas only the
LSMO magnetism is probed at 642.5 eV. The observation of soft
hysteresis loops at all temperatures for Mn at 642.5 eV is direct
evidence that the LSMO is not exchange coupled to the magnetically
harder NMO whose onset of ferrimagnetism is verified at 640 eV. A
comparison of the hysteresis loops taken at 640 and 642.5 eV at 55 K
is further evidence that only a mere superposition of the Mn signals
from the LSMO and NMO takes place, and not a correlated magnetic
coupling effect. The lack of magnetic coupling at the LSMO-NMO
interface likely arises from misfit dislocations due to the lattice
mismatch between the perovskite and spinel structures, providing for
easy 180$^\circ$ domain wall creation at the LSMO-NMO interface, and
decoupling the magnetic layers. This lack of coupling at this
interface enables the independent switching of the two electrodes
below the magnetic ordering temperature of the NMO. This independent
switching is further verified in symmetric minor loops of LSMO.

XMCD hysteresis loops were also taken at various temperatures on the
trilayer sample to probe the magnetic behavior of the
NMO-Fe$_{3}$O$_{4}$ interface. A distinct increase in coercive field
of the Fe$_{3}$O$_{4}$ layer is seen below the NMO T$_{C}$. This
increase, as well as coincident Fe, Mn, and Ni loops, reveals the
presence of strong ferromagnetic coupling between the two spinel
layers, while still maintaining the onset of NMO
ferrimagnetism.\cite{N-C08}

Now that we have established the magnetic decoupling of the
ferromagnetic electrodes despite the magnetic barrier, we focus on
the spin transport in the junction above and below the T$_{C}$ of
the barrier layer. As shown in Fig. 2(a), transitions in the
magnetization hysteresis loops coincide well with large and abrupt
transitions in the JMR. The antiparallel LSMO-Fe$_{3}$O$_{4}$
magnetization configuration is the low resistance state, resulting
in a negative magnetoresistance. This is due to the opposite spin
polarizations of LSMO and Fe$_{3}$O$_{4}$ electrodes, which are
majority and minority spin polarized, respectively.\cite{Hu02}

The temperature dependence of the JMR is studied to investigate the
effect of the barrier layer magnetism on the magnetotransport of the
junctions [Fig. 2(b)]. The maximum JMR increases in magnitude with
decreasing temperature until reaching a maximum around 45 K, and
then decreases with subsequent decrease in temperature. A dramatic
increase in junction resistance is coincident with the decrease in
JMR. A change in JMR bias dependence also occurs as a function of
temperature. At high temperatures, the JMR drops off gradually at
higher voltages and has a skewed, asymmetric behavior in which the
JMR maximum is observed at a finite voltage. A transition occurs
around 60 K, in which the bias dependence transitions to a symmetric
bias dependence, characterized by a JMR maximum at zero bias, a
highly symmetric shape, and a rapid decrease in JMR for V$>$50-100
mV. At low temperatures, the bias dependence continues to be
symmetric about zero bias, although the maximum JMR decreases with
decreasing temperature. At these lower temperatures, the change in
resistance $\triangle$R due to the application of a magnetic field
continues to increase with decreasing temperature despite the fact
that the JMR decreases. Similar behavior is seen for all barrier
layer thicknesses.
\begin{figure}
\center{\includegraphics[width=8 cm]{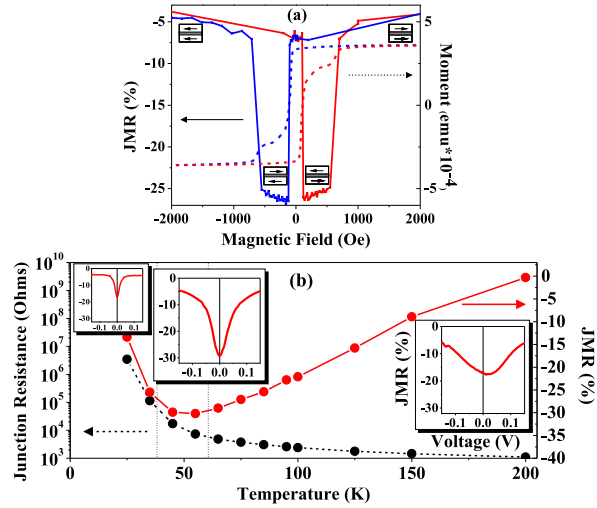}}
\caption{(Color online) (a) Switching characteristics of
Fe$_{3}$O$_{4}$/NMO/LSMO junctions: magnetic hysteresis loop and
resistance vs applied magnetic field at 75 K. (b) Junction
resistance and maximum JMR as a function of temperature, with bias
dependence of JMR for each regime.}
\end{figure}

In order to investigate the inelastic tunneling processes in our
junctions, we plot the second derivative of the IV curves, known as
the IETS. The IETS illustrates the inelastic tunneling processes due
to phonons and spin waves, or magnons. In order to isolate the
effects of magnetism on the inelastic transport, IETS from the two
distinct magnetization configurations, (IETS)$_{P}$ and
(IETS)$_{AP}$, are subtracted from one another. The resulting IETS
(Fig. 3) shows inelastic tunneling events related solely to magnons
as all other factors are held constant between both magnetization
configurations.\cite{Moodera98} The d$^{2}$I/dV$^{2}$=0 value occurs
at a finite bias for T$>$60 K, but moves towards zero bias below 60
K. The peak locations (30-50 mV) of the IETS of our junctions for
all temperatures are well within the range of 12-100 mV commonly
associated with magnons.\cite{Ando00,Tsui71}

The observation of asymmetric bias dependence above the ordering
temperature of NMO and symmetric bias dependence below the ordering
temperature of NMO in both the JMR and IETS suggests that different
conduction mechanisms are dominant depending on the magnetic state
of the barrier. There have been numerous studies of MTJs where
asymmetries in the JMR bias dependence have been attributed to two
different interface density of states at the two electrode-barrier
interfaces. Magnon-assisted tunneling current from electrode surface
magnons has also been predicted to be asymmetric with respect to
bias. Moodera \emph{et al.} proposed that part of the large decrease
in JMR with bias can be attributed to the excitation of magnons with
increasing bias, which randomize the tunneling electron
spins.\cite{Moodera99} Thus an asymmetry in JMR could also be caused
by the unequal number of magnons created at each electrode surface
as a result of the inevitable difference in the two
electrode-barrier interfaces.\cite{Ando00,Moodera98} In our
junctions, the Fe$_{3}$O$_{4}$-NMO interface is isostructural since
both crystallize in the spinel structure; the LSMO-NMO interface is
spinel/cubic perovskite where antiphase boundaries inevitably result
in a more disordered interface. These two different interfaces give
rise to two different barrier heights, two different interface
density of states, and two different magnon density of states. Thus
the asymmetry in the JMR bias dependence can be attributed to a
number of mechanisms, all of which are associated with the
interface.
\begin{figure}
\center{\includegraphics[width=6.5 cm]{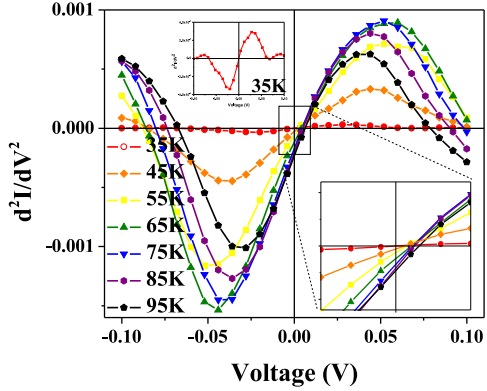}} \caption{(Color
online) Subtracted IETS as a function of temperature, showing shift
towards symmetry about zero bias below 65 K.} \label{IETS}
\end{figure}

However, while the observation of an asymmetric bias dependence due
to the electrode-barrier interfaces is not unusual, the observation
of a symmetric and asymmetric bias dependence of the JMR at
different temperatures in the same junction is a surprising result
and sheds light on the role of the barrier in the tunneling process.
Such changes in symmetry are also observed in the IETS data. The
asymmetric bias dependence of the IETS above the magnetic ordering
temperature of the NMO is indicative of a difference in the forward
and reverse magnon-assisted conduction and is consistent with
transport dominated by electrode-barrier interface magnons. As the
temperature is lowered below the magnetic ordering temperature of
the NMO (60 K), the bias dependence of the JMR and IETS becomes
symmetric, thus suggesting that the dominant conduction mechanism is
no longer sensitive to interface effects but to the bulk of the
barrier. Gajek \emph{et al.} postulated that the symmetric drop of
JMR with bias for their spin filter junctions with ferromagnetic
barrier layers was due to magnon excitations inside the barrier
layer.\cite{Gajek05} Recalling that the subtracted IETS highlights
the inelastic transport due to only magnons in the junction, the
fact that the IETS becomes symmetric with respect to bias at around
60 K suggests that magnons within the barrier are beginning to
influence the spin dependent transport. Such magnons from the bulk
of the barrier would manifest themselves in the same way regardless
of the direction of the current and thus give rise to a symmetric
bias dependence both in the IETS and JMR. Furthermore, a closer look
at the bias dependence of the JMR indicates a dramatic decrease in
the JMR as a function of increasing bias. Below the magnon
excitation energy of about 30-50 mV ascertained from the IETS data,
spin polarized transport is preserved, resulting in significant JMR;
above this voltage, magnons are excited and spin polarized transport
is suppressed.

At even lower temperatures, the JMR bias dependence remains
symmetric but the magnitude of the JMR begins to decrease around 45
K. This decrease is attributed to the dramatic increase in the
junction resistance that is due to the suppressed metal-insulator
transition of the Fe$_{3}$O$_{4}$ electrode, causing the electrode
resistance to rival the barrier resistance.\cite{Alldredge06} As a
function of decreasing temperature, the increase in electrode
resistance along with a comparable $\triangle$R results in an
overall decrease in JMR. Similar junctions with nonmagnetic barriers
also show peaks in the JMR, precluding the possibility that the
decrease in the JMR is related to the magnetic transition of the
barrier.\cite{Alldredge06} The symmetric bias dependence of the JMR
and the increase in $\triangle$R in this temperature range suggests
that the magnetotransport continues to be dominated by bulk magnon
excitations in the NMO barrier layer.

In summary, spin polarized transport in our Fe$_{3}$O$_{4}$/NMO/LSMO
junctions reveals that the barrier plays an important role in the
tunneling process. In these junctions, we have shown that spin
transport is dominated by the interfaces or by the barrier itself
depending on whether NMO is paramagnetic or ferrimagnetic. This
behavior is observed via a distinct change in the bias dependence of
the MTJs which coincides with the onset of ferrimagnetism in the
NiMn$_{2}$O$_{4}$ barrier layer as confirmed by XMCD. When the
barrier layer is paramagnetic at high temperatures, the
electrode-barrier interfaces determine the asymmetric JMR bias
dependence due to their different electrode-barrier surface magnon
properties and different barrier heights. As the temperature is
lowered, the JMR bias dependence becomes highly symmetric,
suggesting a conduction mechanism involving magnons within the
ferrimagnetic NMO barrier layer. Such hybrid behavior in a single
junction is promising for future spintronic applications.

This work was supported in full by the Office of Basic Energy
Sciences, Division of Materials Sciences and Engineering, of the
U.S. Department of Energy under Contract No. DE-AC02-05CH11231.
Thanks to J.S. Moodera for fruitful discussion. Processing was
performed in the University of California - Berkeley Microlab.

\end{document}